\documentclass[
    ,final            
  ]
  {aipproc}

\layoutstyle{8x11double}

\begin{document}

\title{X-ray follow-up observations of unidentified \\ VHE $\gamma$-ray sources
}


\author{Gerd P{\"u}hlhofer}{
  address={Landessternwarte, K{\"o}nigstuhl, 69117 Heidelberg, Germany},
  altaddress={Institut f{\"u}r Astronomie und Astrophysik, Sand 1, 72076 T{\"u}bingen, Germany}
}



\begin{abstract}
A large fraction of the recently discovered Galactic Very High Energy (VHE) source population remains unidentified to date. VHE $\gamma$-ray emission traces high energy particles in these sources, but for example in case of hadronic processes also the gas density at the emission site. Moreover, the particles have sufficiently long lifetimes to be able to escape from their acceleration sites. Therefore, the $\gamma$-ray sources or at least the areas of maximum surface brightness are in many cases spatially offset from the actual accelerators. A promising way to identify the objects in which the particles are accelerated seems to be to search for emission signatures of the acceleration process (like emission from shock-heated plasma). Also the particles themselves (through primary or secondary synchrotron emission) can be traced in lower wavebands. Those signatures are best visible in the X-ray band, and current X-ray observatories are well suited to conduct such follow-up observations. Some aspects of the current status of these investigations are reviewed.
%
%
%
\end{abstract}

\maketitle


\section{Introduction}

\begin{figure}
  \includegraphics[width=.98\textwidth]{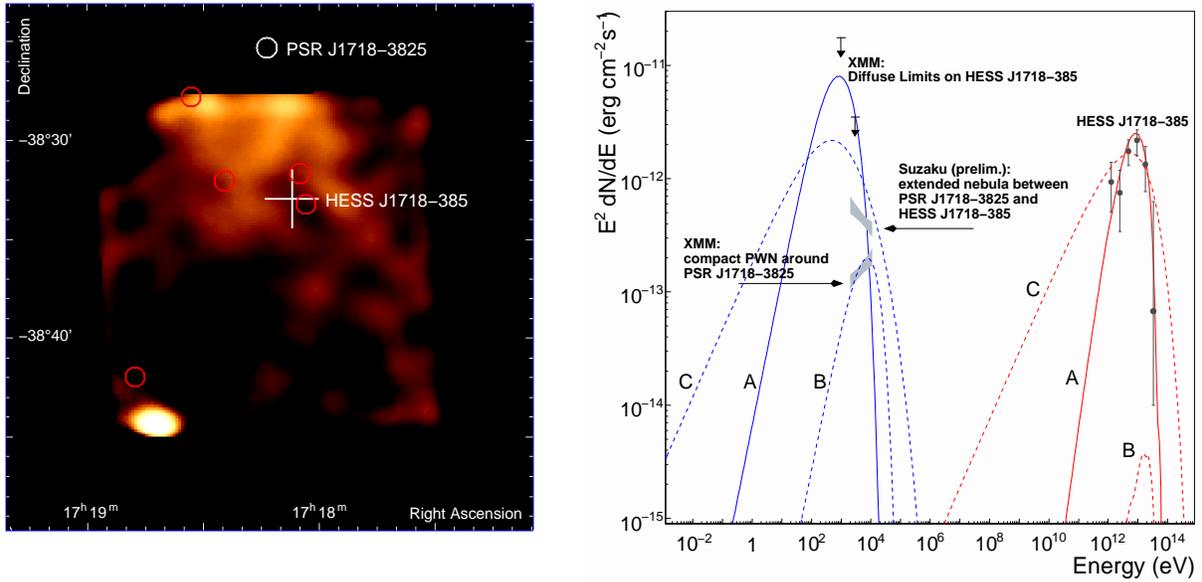}
  \caption{Results from X-ray observations of the VHE PWN candidate HESS\,J1718-385. Left: Image obtained with the Suzaku XIS\,0 detector in the 0.5-7\,keV band, in a 15\,ksec exposure. The colour scale was adjusted to emphasize the diffuse structure seen towards the North of the image. The white cross indicates the position of the VHE source centroid. Red circles indicate the position of point sources detected with XMM-Newton. The excess in the Northern half of the image -- compared to a mirrored control region in the Southern part of the image -- is significant. The source can be interpreted as a low surface brightness X-ray PWN, which connects the compact PWN around PSR\,J1718-3825 as detected with XMM-Newton \cite{Hinton20081718} with the VHE source. The XIS spectrum (derived from all three operational XIS detectors) is compatible with a power law with photon index $\Gamma=2.1^{+0.10}_{-0.16}$; the systematic error from the background selection and uncertainty of the absorption column is estimated to $\pm 0.3$. Right: Spectral energy distribution of the HESS\,J1718-385 area. The image is adopted from \cite{Hinton20081718}. Models A and B represent a hypothetical very narrow electron injection spectrum close to the pulsar (B) and after propagation to match the extension of the VHE source (A). Model C serves to illustrate that a more conventional, power-law electron distribution can also be invoked to fit the data. The Suzaku XIS spectrum fits to the expectations of a cooling electron distribution, injected by the pulsar PSR\,J1718-3825.}
  \label{F:Puehlhofer_Fig1}
\end{figure}

In the last couple of years, a considerable number of new VHE $\gamma$-ray sources was discovered in the Galactic plane. Many of these sources were discovered in survey observations performed with the H.E.S.S. array of Cherenkov telescopes \cite{HESS2005SurveyI,HESS2006SurveyII,HESS2007twoPWN,HESS2008Darksources}, or in the field of view of H.E.S.S. observations of other targets in the Galactic plane (e.g. \cite{HESS20051303, HESS2007Monoceros}). A considerable fraction of the new sources could not be identified with known astrophysical objects as identified in lower wavebands. As a consequence, follow-up observations especially in the X-ray band were performed with the aim to identify the objects that accelerate particles to very high energies visible in VHE $\gamma$-rays \cite{Puehlhofer2008Xray}.

The paper focuses on two topics which arose during the identification efforts using X-ray telescopes: Pulsar wind nebulae (PWN), especially their appearance as offset, {\em relic} PWN, and the investigation of VHE emission sites coinciding with molecular clouds.

\section{Relic Pulsar Wind Nebulae}

The apparent (physical and angular) size of a PWN is given by the livetime of the electrons that emit in the respective frequency band, in which the source is observed. In general, one expects the most compact appearance in synchrotron X-rays, a larger size in Inverse Compton (IC) VHE $\gamma$-rays, and the biggest extension in radio synchrotron emission. For the ``classical'' PWN, the Crab nebula, however, the point spread function of present VHE telescopes of $\sim 0.1^{\circ}$ does not allow to resolve the VHE nebula. 

The flux ratio between X-rays and VHE $\gamma$-rays, $F_{\mathrm{X}}/F_{\mathrm{VHE}}$, in the Crab nebula is $\sim 100$. It is however expected that pulsars older than the Crab pulsar (e.g. ``middle-aged'' pulsars) exhibit larger IC nebulae of electrons that have escaped the high B-field region close to the pulsar, and could accumulate over a significant fraction of the pulsar's livetime \cite{Aharonian1997PWN}. In this case, flux ratios as low as $F_{\mathrm{X}}/F_{\mathrm{VHE}} \sim 10 .. 1$ are expected and observed, e.g. in G\,0.9-0.1, MSH\,15-5{\it 2}, or Vela X. 

It turns out that some of the unidentified VHE sources can be explained in a relic PWN scenario, with flux ratios $F_{\mathrm{X}}/F_{\mathrm{VHE}}$ even smaller than 1 ($\sim 0.1$). This was exemplified in the case of HESS\,J1640-465, where an X-ray PWN candidate discovered with XMM-Newton could be successfully identified with the positionally coincident VHE source \cite{Funk20071640}.
Because of the potential faintness of the X-ray counterparts to the VHE sources, and because of the large absorption column towards many of the VHE sources, it is likely that previous X-ray surveys with ROSAT and ASCA have missed those sources. 
Flux ratios as low as $F_{\mathrm{X}}/F_{\mathrm{VHE}} \sim 0.01$ are meanwhile considered plausible, because the synchrotron SED peak that corresponds to the IC VHE peak may be located in the UV band in case of low magnetic fields\footnote{ 
One should note that $F_{\mathrm{X}}/F_{\mathrm{VHE}}$ depends on the choice of the X-ray and VHE $\gamma$-ray band, and that fluxes used to compute $F_{\mathrm{X}}/F_{\mathrm{VHE}}$ are not necessarily derived from sky areas of the same angular size, and are hence probing different volumes in space.}.

In addition to the different angular scales, VHE and X-ray PWN can be located spatially offset from the powering pulsar, e.g. due to an inhomogeneous surrounding medium in a crushed PWN scenario \cite{Blondin2001PWN}, or because of a high velocity of the pulsar. In such a case, the VHE and X-ray PWN centroids are both offset from the pulsar (in the same angular direction), and moreover the VHE centroid can be more displaced than the X-ray centroid. Such a scenario has been demonstrated in the case of the proven association between 
HESS\,J1825-137 and G\,18.0-0.7 (\cite{HESS20051825,HESS20061825})\footnote{The association in this case was proven by a softening of the VHE spectrum away from the pulsar, in addition to the common offset direction of the X-ray and VHE centroid and a plausible examination of the pulsar and particle energetics.}. 

Using the hypothesis of an offset, relic PWN scenario, some VHE sources were classified as VHE PWN candidates, based on the proximity of an energetic pulsar in reasonable angular distance to the VHE source (e.g. \cite{Gallant2006PWN,HESS2007twoPWN}). Follow-up observations with X-ray telescopes were performed with the aim to identify the previously undetected X-ray PWN around the pulsar and to search for asymmetries that would confirm the association with the offset VHE source. For example, follow-up observations of HESS\,J1718-385 with XMM-Newton indeed identifed an X-ray PWN around PSR\,J1718-3825 \cite{Hinton20081718}. The morphological connection could not be proven with the XMM-Newton data alone; however, a Suzaku observation of the same area indicates that indeed a weak diffuse nebula connects PSR\,J1718-3825 and the VHE centroid of HESS\,J1718-385, see Fig.\ref{F:Puehlhofer_Fig1}.

\section{Pulsar Wind Nebula Pevatrons?}

Some pulsars that are -- confirmed or likely -- connected to VHE sources via a PWN scenario also exhibit hard X-ray emission above 10\,keV, as detected with BeppoSAX and INTEGRAL (e.g. \cite{Dean20081813Integral,Hoffmann2006Integral}). The hard X-ray emission can either be explained by magnetospheric or PWN emission, or a superposition of both. To disentangle the two possible components, one can try to separate the pulsed and unpulsed fraction of the flux, and/or use the limited imaging capabilites of the IBIS/ISGRI detector onboard INTEGRAL (see next section). 

If the X-ray PWN emission dominates over the nonthermal pulsar in the 1-10\,keV range, where a separation is possible using current X-ray imaging instruments, then also a spectral analysis can be employed to test whether the PWN emission below 10\,keV and the spatially unresolved emission above 10\,keV is plausibly connected. 

In the case of HESS\,J1813-178, the X-ray PWN candidate discovered with XMM-Newton (and also found in archival ASCA data) is spectrally well connected to the spatially coincident INTEGRAL source discovered by \cite{Ubertini20051813Integral}. \cite{Funk20071813} showed that in a PWN scenario, such an identification indicates the acceleration of high energy particles to PeV energies. However, the interpretation in \cite{Funk20071813} relied yet on the assumption that the X-ray emission is dominated by PWN emission, and not by the putative pulsar unresolved in XMM-Newton. A subsequent Chandra observation \cite{Helfand20071813} could confirm this assumption, by spatially identifying the pulsar candidate and showing that the PWN dominates over the oulsar emission in the $<$10\,keV band by a factor of $\sim 5$. Such a value is typical for PWNe powered by energetic pulsars \cite{Kargaltsev2007Chandrareview,Gotthelf2003PWN}.

It should be noted that an alternative scenario of particles accelerated in the Supernova remnant (SNR) shell, as seen in radio observations \cite{Brogan20051813}, might also explain the VHE emission in HESS\,J1813-178 \cite{Funk20071813}. The angular re\-solution of the VHE data does not allow to disentangle the PWN and the SNR shell scenario in this case \cite{Funk20071813} .

\section{``Underluminous'' X-ray Pulsar Wind Nebulae}

The efficiencies with which the spin-down luminosity of a pulsar is converted into the X-ray PWN ($L_{\mathrm{X, PWN}}$) and into nonthermal magnetospheric emission ($L_{\mathrm{X, PSR}}$) exhibit a large scatter. On the other hand, both luminosities are fairly tightly correlated, with $L_{\mathrm{X, PWN}} \sim 5 \times L_{\mathrm{X, PSR}}$, as shown e.g. by \cite{Kargaltsev2007Chandrareview} in a Chandra study of energetic pulsars. There are, however, some noteworthy exceptions, where the X-ray PWN is considerably fainter than the pulsar (``underluminous'' PWN, \cite{Kargaltsev20081617}). Of the three underluminous sources with $L_{\mathrm{PWN, 0.5-8\,keV}} > 10^{33}\mathrm{erg\,s^{-1}}$, one is a confirmed VHE PWN emitter (MSH\,15-5{\it 2}/PSR\,B1509-58 \cite{HESS2005MSH1552}), whereas the other two pulsar/PWN systems are currently being discussed as possible counterparts to VHE sources: The newly discovered pulsar PSR\,J1838-0655 with HESS\,J1837-069 \cite{Gotthelf20081837,Marandon20081837}, and PSR\,J1617-5055 with HESS\,J1616-508 \cite{Landi20071616,Kargaltsev20081617}. 

All three sources are also detected in the hard X-ray band above 10\,keV with INTEGRAL and BeppoSAX. The sources are therefore of interest because they could prove to be further pevatron PWN systems.

{\it MSH\,15-5{\it 2}/PSR\,B1509-58}: Indeed, using INTEGRAL imaging \cite{Forot2006msh1552} could show that the hard unpulsed X-ray source is likely driven by the PWN, and derive cutoff energies of 0.4..0.7\,PeV. The identification of the VHE source with the X-ray PWN was already demonstrated in \cite{HESS2005MSH1552}.

{\it PSR\,J1617-5055/HESS\,J1616-508}: Here, the situation is unresolved. Concerning the identification of the hard X-ray source with the soft X-ray PWN, it looks like the INTEGRAL/BeppoSAX source is dominated by the pulsar itself, there is no evidence for hard X-ray PWN emission yet. Since the X-ray PWN is underluminous \cite{Kargaltsev20081617}, no spectral connection arguments can be made as in the case of HESS\,J1813-178 (see previous section). Nevertheless, \cite{Landi20071616} argued that PSR\,J1617-5055 and the corresponding hard X-ray source is the most likely counterpart to HESS\,J1616-508. But the underluminous PWN does not morphologically connect to the VHE source at all, therefore e.g. \cite{Kargaltsev20081617} pointed out that the PWN driven by PSR\,J1617-5055 might only explain a fraction of the VHE emission. Using an upper limit from Suzaku below 10\,keV, \cite{Matsumoto20071616} even argued that the VHE source could not be connected to PSR\,J1617-5055 at all.

{\it PSR\,J1838-0655/HESS\,J1837-069}: The discovery of a new X-ray pulsar using RXTE coincident with the unresolved source AX\,J1838.0-0655, together with the detection of a faint X-ray nebula around the source with Chandra, was accompanied by the suggestion that this source could be identified with HESS\,J1837-069 in a PWN scenario \cite{Gotthelf20081837,Marandon20081837}. The possible association of the VHE to the ASCA source had already been discussed in \cite{HESS2006SurveyII,Malizia20051838}. However, the hard X-ray source detected with INTEGRAL \cite{Lutovinov2005IntegralHMXB,Malizia20051838} spectrally connects to the X-ray pulsar rather than to the nebula, and there is no evidence for a $>$10\,keV nebula yet. Again, the faintness of the $<$10\,keV nebula does not allow to argue in favour of a $>$10\,keV PWN as in the case of HESS\,J1813-178. Moreover, the nebula around PSR\,J1838-0655 does not exhibit any asymmetry that would connect the X-ray to the VHE source. Last but not least, the nebula could (at least in part) also be explained by a scattering halo rather than by a PWN, due to the large column density towards the source of $N_{\mathrm{H}}\sim 5 \times 10^{22}\mathrm{cm^{-2}}$ \cite{Schwemmer20081837}.

To summarize, the fact that the X-ray nebulae around PSR\,J1617-5055 and PSR\,J1838-0655 are very faint hampers the identification of the closeby VHE sources in a PWN scenario. Because of the large absorption columns of $3.5 .. 5 \times 10^{22}\mathrm{cm^{-2}}$, detections below 2\,keV are not possible, preventing e.g. earlier ROSAT detections of putative soft and extended nebulae. Both pulsars also have another pulsar (PSR\,B1610-50) and possible pulsar/PWN system (AX\,J1837.3-0652) nearby, which could contribute to the VHE sources. Hence, a unique identification of both the VHE and the hard X-ray sources with the corresponding X-ray PWNe is debatable, as discussed above. The interpretation of both systems as pevatron PWNe is also not conclusively possible at this stage.

\section{VHE sources coincident with molecular clouds interacting with SNR shells}

Two VHE sources were discovered recently which coincide with molecular clouds that may have been shocked by SNR shocks, as indicated by high magnetic fields (0.2..0.6\,mG) observed in Zeeman splitting of maser lines: HESS\,J1745-303, or more specifically its region A which is coincident with part of the shell of G\,359.3-0.5 \cite{HESS20081745}, and HESS\,J1714-385 coincident with the SNR CTB\,37A \cite{HESS2008CTB37A}. To explain the VHE sources by IC emission of energetic electrons would imply intense fluxes of X-ray synchrotron emission. Therefore, and because of the high gas densities in the clouds ($10^2 .. 10^3\mathrm{cm}^{-3}$), the presence of high energy hadrons is the most likely scenario to explain the VHE emission in both sources. 

X-ray observations of both VHE sources have not yielded evidence for diffuse non-thermal X-ray emission in spatial agreement with the VHE sources, strengthening the case for a hadronic scenario. However, in CTB\,37A, an X-ray PWN candidate coincident with the VHE source was discovered in Chandra observations \cite{HESS2008CTB37A}. Since this source might only in projection be coincident with the molecular cloud and could therefore be located in a low B-field environment, the alternative scenario of a relic PWN to explain the VHE emission from HESS\,J1714-385 can not be excluded conclusively yet.

Even in a hadronic scenario, synchrotron emission from electrons produced in hadronic collisions is inevitably expected. Current limits on the diffuse X-ray component of these sources are getting close to the expectations.

\begin{theacknowledgments}
G.P. acknowledges support by the German Ministry for Education and Research (BMBF) through DESY grant  05\,CH5\,PC1/6 and  DLR grant 50\,OR\,0502. 
\end{theacknowledgments}



\bibliographystyle{aipproc}   

\bibliography{sample}


\end{document}